\documentclass[aps,pra,reprint,superscriptaddress,nofootinbib]{revtex4-2}
\usepackage{graphicx}
\usepackage{subfigure}
\usepackage{dcolumn}
\usepackage{hyperref}
\usepackage{amsmath}

\usepackage{orcidlink}

\begin{document}

%\preprint{}

\title{Single and double \textit{K}-shell vacancy production in slow Xe$^\textrm{54+,53+}$-Xe collisions}

\author{P.-M.~Hillenbrand\,\orcidlink{0000-0003-0166-2666}}
\affiliation{Institut f\"ur Kernphysik, Goethe-Universit\"at, 60438 Frankfurt, Germany}
\affiliation{GSI Helmholtzzentrum f\"ur Schwerionenforschung, 64291 Darmstadt, Germany}

\author{S.~Hagmann\,\orcidlink{0000-0001-8079-3546}}
\affiliation{GSI Helmholtzzentrum f\"ur Schwerionenforschung, 64291 Darmstadt, Germany}

\author{Y.~S.~Kozhedub\,\orcidlink{0000-0003-1273-9008}}
\affiliation{Department of Physics, St.~Petersburg State University, 199034 St.~Petersburg, Russia}

\author{E.~P.~Benis\,\orcidlink{0000-0002-5564-153X}}
\affiliation{Department of Physics, University of Ioannina, 45110 Ioannina, Greece}

\author{C.~Brandau\,\orcidlink{0000-0001-8825-8820}}
\affiliation{GSI Helmholtzzentrum f\"ur Schwerionenforschung, 64291 Darmstadt, Germany}
\affiliation{I.~Physikalisches Institut, Justus-Liebig-Universit\"at, 35392 Giessen, Germany}

\author{R.~J.~Chen}
\affiliation{GSI Helmholtzzentrum f\"ur Schwerionenforschung, 64291 Darmstadt, Germany}

\author{D.~Dmytriiev\,\orcidlink{0000-0003-3612-2686}}
\affiliation{GSI Helmholtzzentrum f\"ur Schwerionenforschung, 64291 Darmstadt, Germany}
\affiliation{Fakult\"at f\"ur Physik und Astronomie, Ruprecht-Karls-Universit\"at, 69117 Heidelberg, Germany}

\author{O.~Forstner\,\orcidlink{0000-0003-3636-4669}}
\affiliation{GSI Helmholtzzentrum f\"ur Schwerionenforschung, 64291 Darmstadt, Germany}
\affiliation{Institut f\"ur Optik und Quantenelektronik, Friedrich-Schiller-Universit\"at, 07743 Jena, Germany}

\author{J.~Glorius\,\orcidlink{0000-0003-1785-3221}}
\affiliation{GSI Helmholtzzentrum f\"ur Schwerionenforschung, 64291 Darmstadt, Germany}

\author{R.~E.~Grisenti\,\orcidlink{0000-0002-9895-6558}}
\affiliation{Institut f\"ur Kernphysik, Goethe-Universit\"at, 60438 Frankfurt, Germany}
\affiliation{GSI Helmholtzzentrum f\"ur Schwerionenforschung, 64291 Darmstadt, Germany}

\author{A.~Gumberidze\,\orcidlink{0000-0002-2498-971X}}
\affiliation{GSI Helmholtzzentrum f\"ur Schwerionenforschung, 64291 Darmstadt, Germany}

\author{M.~Lestinsky\,\orcidlink{0000-0001-7384-1917}}
\affiliation{GSI Helmholtzzentrum f\"ur Schwerionenforschung, 64291 Darmstadt, Germany}

\author{Yu.~A.~Litvinov\,\orcidlink{0000-0002-7043-4993}}
\affiliation{GSI Helmholtzzentrum f\"ur Schwerionenforschung, 64291 Darmstadt, Germany}
\affiliation{Fakult\"at f\"ur Physik und Astronomie, Ruprecht-Karls-Universit\"at, 69117 Heidelberg, Germany}

\author{E.~B.~Menz\,\orcidlink{0000-0003-1451-7089}}
\affiliation{GSI Helmholtzzentrum f\"ur Schwerionenforschung, 64291 Darmstadt, Germany}
\affiliation{Institut f\"ur Optik und Quantenelektronik, Friedrich-Schiller-Universit\"at, 07743 Jena, Germany}
\affiliation{Helmholtz-Institut Jena, 07743 Jena, Germany}

\author{T.~Morgenroth}
\affiliation{GSI Helmholtzzentrum f\"ur Schwerionenforschung, 64291 Darmstadt, Germany}
\affiliation{Institut f\"ur Optik und Quantenelektronik, Friedrich-Schiller-Universit\"at, 07743 Jena, Germany}
\affiliation{Helmholtz-Institut Jena, 07743 Jena, Germany}

\author{S.~Nanos\,\orcidlink{0000-0001-5428-3518}}
\affiliation{Tandem Accelerator Laboratory, INPP, NCSR “Demokritos”, 15310 Agia Paraskevi, Greece}
\affiliation{Department of Physics, University of Ioannina, 45110 Ioannina, Greece}

\author{N.~Petridis}
\affiliation{GSI Helmholtzzentrum f\"ur Schwerionenforschung, 64291 Darmstadt, Germany}

\author{Ph.~Pf\"afflein\,\orcidlink{0000-0003-0517-0722}}
\affiliation{GSI Helmholtzzentrum f\"ur Schwerionenforschung, 64291 Darmstadt, Germany}
\affiliation{Institut f\"ur Optik und Quantenelektronik, Friedrich-Schiller-Universit\"at, 07743 Jena, Germany}
\affiliation{Helmholtz-Institut Jena, 07743 Jena, Germany}

\author{H.~Rothard}
\affiliation{Centre de Recherche sur les Ions, les Mat\'eriaux et la Photonique CIMAP, Normandie Universit\'e, ENSICAEN, UNICAEN, CEA, CNRS, 14000 Caen, France}

\author{M.~S.~Sanjari\,\orcidlink{0000-0001-7321-0429}}
\affiliation{GSI Helmholtzzentrum f\"ur Schwerionenforschung, 64291 Darmstadt, Germany}
\affiliation{Aachen University of Applied Sciences, 52066 Aachen, Germany}

\author{R.~S.~Sidhu\,\orcidlink{0000-0002-1637-7502}}
\affiliation{GSI Helmholtzzentrum f\"ur Schwerionenforschung, 64291 Darmstadt, Germany}
\affiliation{Fakult\"at f\"ur Physik und Astronomie, Ruprecht-Karls-Universit\"at, 69117 Heidelberg, Germany}

\author{U.~Spillmann\,\orcidlink{0000-0001-7281-5063}}
\affiliation{GSI Helmholtzzentrum f\"ur Schwerionenforschung, 64291 Darmstadt, Germany}

\author{S.~Trotsenko}
\affiliation{GSI Helmholtzzentrum f\"ur Schwerionenforschung, 64291 Darmstadt, Germany}

\author{I.~I.~Tupitsyn\,\orcidlink{0000-0001-9237-5667}}
\affiliation{Department of Physics, St.~Petersburg State University, 199034 St.~Petersburg, Russia}

\author{L.~Varga}
\affiliation{GSI Helmholtzzentrum f\"ur Schwerionenforschung, 64291 Darmstadt, Germany}
\affiliation{Fakult\"at f\"ur Physik und Astronomie, Ruprecht-Karls-Universit\"at, 69117 Heidelberg, Germany}

\author{Th.~St\"ohlker\,\orcidlink{0000-0003-0461-3560}}
\affiliation{GSI Helmholtzzentrum f\"ur Schwerionenforschung, 64291 Darmstadt, Germany}
\affiliation{Institut f\"ur Optik und Quantenelektronik, Friedrich-Schiller-Universit\"at, 07743 Jena, Germany}
\affiliation{Helmholtz-Institut Jena, 07743 Jena, Germany}

\date{\today}

\begin{abstract}
We present an experimental and theoretical study of symmetric $\textrm{Xe}^{54+}+\textrm{Xe}$ collisions at 50, 30, and 15~MeV/u, corresponding to strong perturbations with $v_K/v_\text{p}=1.20$, 1.55, and 2.20, respectively ($v_K$: classical $K$-shell orbital velocity, $v_\text{p}$: projectile velocity), as well as   $\textrm{Xe}^{53+}+\textrm{Xe}$ collisions at 15~MeV/u. For each of these systems, x-ray spectra were measured under a forward angle of $35^\circ$ with respect to the projectile beam. Target satellite and  hypersatellite radiation, $K\alpha_{2,1}^\mathrm{s}$ and $K\alpha_{2,1}^\mathrm{hs}$, respectively, were analyzed and used to derive cross section ratios for double-to-single target $K$-shell vacancy production. We compare our experimental results to  relativistic time-dependent two-center calculations.
\end{abstract}

\maketitle

\section{Introduction}

Due to its nonperturbative character, the accurate description for the propagation of bound electrons in the field of two heavy nuclei colliding at slow velocities is a fundamental challenge on the path to understanding heavy-ion collision dynamics \cite{bates_inelastic_1953}. In these systems, the collision partners can be described by transiently forming heavy quasimolecules, as the atomic orbitals merge into molecular orbitals that evolve as a function of the impact parameter and the internuclear distance \cite{bosch_quasimolecular_1980}. Studies on slow collisions of heavy ions are strongly motivated by the prediction of spontaneous electron-positron pair creation when the combined nuclear charge of target and projectile reaches a critical value of $Z_\text{cr}\approx173$ \cite{reinhardt_quantum_1977}. Favorable collision energies corresponding to long ``lifetimes'' of the transient quasimolecule are comparable to the Coulomb barrier and amount to a few MeV/u \cite{heiss_quantum_1985}. These theoretical predictions have recently been refined \cite{maltsev_electron-positron_2015,maltsev_electron-positron_2018,maltsev_how_2019,popov_how_2020}, based on major advances in the relativistic time-dependent two-center description of the process \cite{tupitsyn_relativistic_2010,tupitsyn_relativistic_2012}. 

Numerous experiments with slow heavy ions at particle accelerators focused on impact-parameter dependencies of charge transfer processes \cite{bosch_quasimolecular_1980,hagmann_quasiresonant_1982,hagmann_quasi-resonant_1983,hagmann_k-k_1987, schuch_k-shell_1983,warczak_k_1983, mokler_vacancy_1984, schuch_interference_1984,schuch_quasimolecular_1988,schuch_quantum_2020,schulz_k-shell_1987}.  For incoming bare projectiles, single and double $K$-shell vacancy production was investigated, which for near-symmetric systems towards low collision energies is dominated by quasi-resonant electron transfer from the target $K$-shell to the projectile $K$-shell \cite{schulz_k-shell_1987,hall_double_1981,wohrer_k-k_1984,hall_energy_1986}. However, fully symmetric systems leading to resonant $K$-shell--to-$K$-shell transfer could not be studied in these experiments, since target and projectile $K$-radiation of identical collision partners could not be resolved at small Doppler shifts. 

In order to produce intense beams of bare or hydrogen-like ions in a stripper target, the beam needs to be accelerated to a velocity larger than the orbital velocity of the projectile $K$-shell electron, $v_\text{p}>v_K$. However, quasi-molecular effects become important at strong perturbations with $v_\text{p}<v_K$. Therefore, early experiments already applied an acceleration-deceleration technique at linear accelerators, where a stripper target was used in between the acceleration and the deceleration stages. Nevertheless, the beam energies available at these facilities limited the projectile charge $q$ and atomic number $Z$ to the mid-$q$ high-$Z$ range \cite{bosch_quasimolecular_1980,mokler_vacancy_1984,schuch_k-shell_1983,warczak_k_1983} or the high-$q$ mid-$Z$ range \cite{schuch_interference_1984,schulz_k-shell_1987,schuch_quasimolecular_1988,schuch_quantum_2020}. 

The development of heavy-ion synchrotrons in combination with cooler-storage rings such as the Experimental Storage Ring (ESR) at the GSI Helmholtzzentrum f\"ur Schwerionenforschung in Darmstadt, Germany, now provides the capabilities to use the driver accelerator to reach the ion energy required for producing the desired charge state through traversing a stripper target, and subsequently use the storage ring to decelerate those ions to the desired low collision velocities, thus providing for the first time highly luminous cooled beams of bare and H-like heavy ions at $v_\mathrm{p}\ll v_K$. First experiments at storage rings with xenon as internal-target gas  have been performed at collision energies down to 50~MeV/u, corresponding to $v_K/v_\text{p}=1.20$ \cite{kozhedub_relativistic_2014,shao_production_2017,gumberidze_impact_2017,yang_alignment_2020, yang_state-selective_2021}. The experimental challenges that need to be overcome when going to even lower collision energies are discussed in this paper. 

In its nonrelativistic straight-line collision description, the impact-parameter dependent probability for single $K$-electron transfer in symmetric collisions of a bare and a hydrogen-like ion can be directly scaled from the $\textrm{H}^+ + \textrm{H}(1s)$ system \cite{winter_electron_2009,abdurakhmanov_accurate_2016}. For heavy ions, relativistic effects are predicted to cause a deviation from this elementary scaling \cite{tupitsyn_relativistic_2010}. In contrast, probabilities for double $K$-electron transfer cannot be scaled directly from the $\textrm{He}^{2+} + \textrm{He}(1s^2)$ system due to the electron-electron interaction and the resulting partially screened Coulomb potential \cite{fritsch_theoretical_1994,baxter_time-dependent_2016}. However, for heavy ions the effectiveness of the electron-electron interaction relative to the Coulomb field strength of the nucleus diminishes, and charge-transfer processes can be readily described in an independent particle model \cite{tupitsyn_relativistic_2010,tupitsyn_relativistic_2012}. Therefore, studies with heavy ions are mainly sensitive to relativistic effects.

The focus of the experimental and theoretical study presented in this paper is the target single and double $K$-shell vacancy production in slow symmetric collisions of bare Xe$^{54+}$ with atomic Xe.  For bare projectiles towards the high collision-energy limit of weak perturbations, target $K$-shell vacancy production is dominated by direct ionization into the continuum. Towards the low collision-energy limit of strong perturbations the adiabatic electron transfer from the target $K$-shell into the projectile $K$-shell proceeding through the transiently formed quasimolecule determines largely the target $K$-shell vacancy production. A resulting distinct signature of the quasi-molecular character is the pronounced probability of double $K$-shell electron transfer investigated in the present work. We present new measurements for bare Xe$^{54+}$ projectiles at 30 and 15~MeV/u, and for H-like Xe$^{53+}$ at 15~MeV/u. Furthermore, we revaluate previous measurements at 50~MeV/u \cite{gumberidze_impact_2017}.

The paper is organized as follows: Sec.~\ref{sec:experiment} provides a brief description of the experiment, Sec.~\ref{sec:results} presents experimental results and their analysis, Sec.~\ref{sec:theory} summarizes the theoretical description of the process, and Sec.~\ref{sec:discussion} discusses the results.

\section{Experiment}\label{sec:experiment}

In the experiment, $^{124}$Xe$^{48+}$ ions were accelerated to 76.5~MeV/u ($v_K/v_\text{p}=0.97$) in the SIS18 synchrotron of GSI. Subsequently, the ions traversed an 11~mg/cm$^2$ carbon stripper target. Xenon ions of the desired charge state, 53+ or 54+, were injected into the ESR storage ring and decelerated to the investigated collision energies. Electron cooling was applied after injection as well as after deceleration. A typical number of ions circulating in the ring at the beginning of the measurement phase was $5\times10^7$. A measurement cycle, e.g.~for Xe$^{54+}$ at 30~MeV/u, comprised an injection every 50~s with a measurement phase of 10 to 20~s. During this measurement phase,
the xenon gas-jet target was activated with an area density of the order of $2\times10^{10}~\mathrm{cm}^{-2}$, a value that was much lower compared to previous similar experiments \cite{kozhedub_relativistic_2014,gumberidze_impact_2017}. Due to the immense cross sections for charge transfer at low collision energies, a stable target operation at this low density was key to the success of the measurements in order to ensure single-collision conditions, acceptable detector count rates, and reasonable beam-storage lifetimes in the ESR.

%The large probabilities for multi-electron capture from the target atom into bound states of the projectile ion could be observed in a position sensitive particle detector mounted for diagnostic purpose in the dispersive dipole section downstream from the gas-jet target. This detector was positioned such that it could resolve projectile ions which had captured between three and seven electrons during the interaction with the target.

X-ray spectra were measured at an angle of $35^\circ$ with respect to the projectile beam direction using an ORTEC GLP series planar high-purity germanium detector with an active diameter of  16~mm. The detector resolution at 30~keV was $\Delta E_\mathrm{FWHM}=357$~eV, as derived from the calibration measurement with an $^{241}$Am x-ray source. A 7-mm slit was used in front of the detector to reduce Doppler-broadening for x rays emitted by the projectile. The distance of the x-ray detector from the interaction point was 34~cm thus resulting in an effective solid angle of $\Delta \Omega/\Omega=7.7\times10^{-5}$. In symmetric collision systems, such as the one studied here, target and projectile radiation can be energetically separated solely by the Doppler shift. For our case of a $35^\circ$ observation angle, this increases the detected energy of x-rays emitted by the projectile by a factor of 1.28, 1.22, and 1.15 at 50, 30, and 15~MeV/u, respectively.

\section{Results}\label{sec:results}

\begin{figure}[b!]
	\centering
	\includegraphics[width=1\columnwidth]{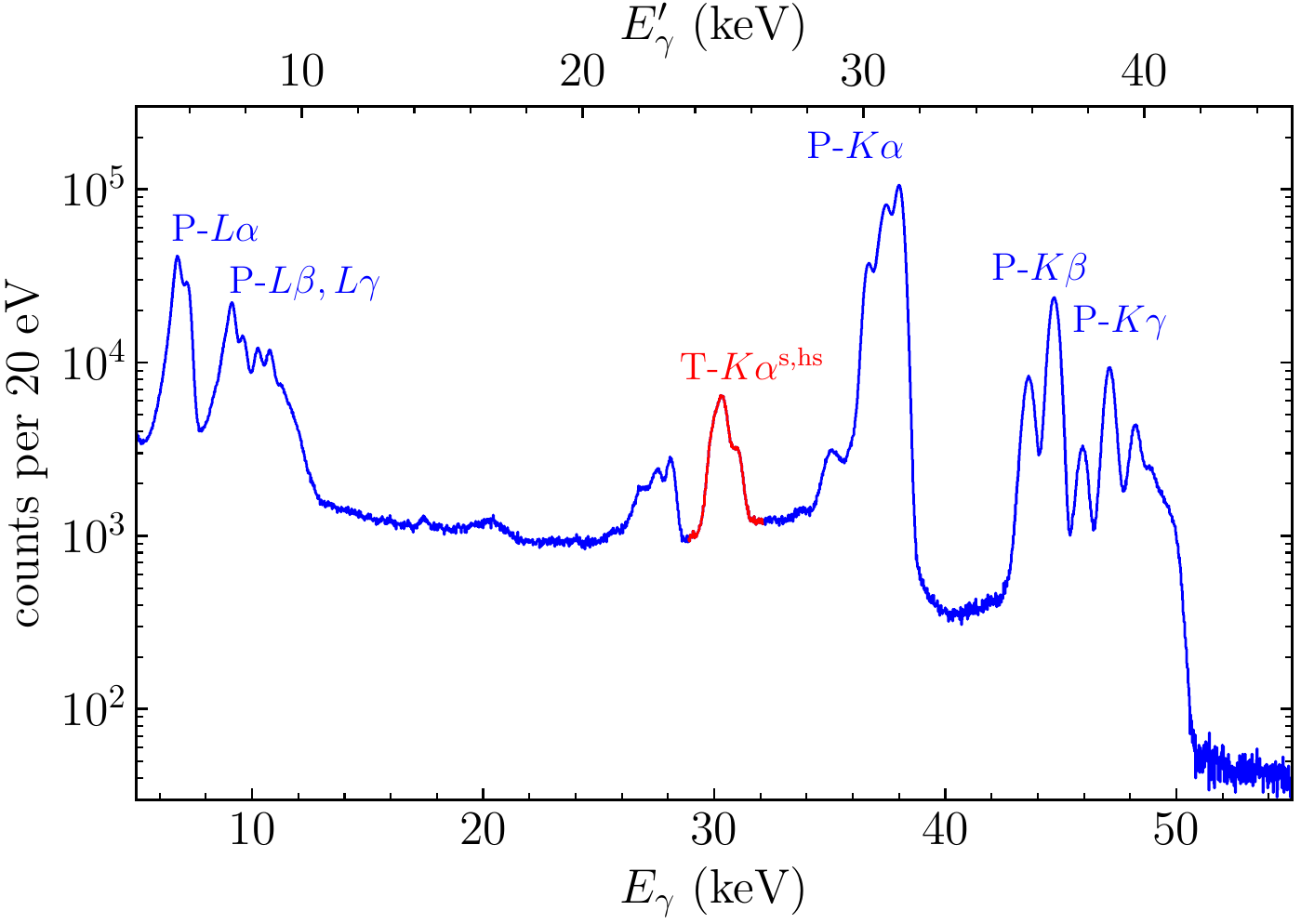}
	\caption{\label{fig:xrayall}X-ray spectrum for Xe$^{54+}+$Xe at 30~MeV/u as a function of the photon energy in the target and the projectile frame, corresponding to the bottom and the top axes, respectively. The unlabeled signal at $\approx 28$~keV is the escape peak.}
\end{figure}

\subsection{Full x-ray spectrum}

The full x-ray spectrum measured for Xe$^{54+}$ projectile at 30~MeV/u is shown in Fig.~\ref{fig:xrayall}. All characteristic features of the projectile (P) radiation after capture of a target electron into an $n\geq2$ state are clearly visible.\footnote{The escape peak originates from events where the detector does not absorb the complete photon energy $E_\gamma$. If one quantum of energy corresponding to a $K\alpha$ transition in the Ge-absorber escapes the detector, the remaining deposited energy causing the escape peak is $E_\mathrm{esc}=E_\gamma-E_{K\alpha}^\mathrm{Ge}\approx E_\gamma-9.9$~keV.} Moreover, the shape of the $\mathrm{P}\mbox{-}K\alpha$ radiation indicates that there is a distinct probability of capturing two electrons leading to a He-like projectile state. Due to the Doppler shift, the $\mathrm{P}\mbox{-}K\alpha$ radiation is well separated energetically from the target (T) satellite (s) and hypersatellite (hs) radiation, $\mathrm{T}\mbox{-}K\alpha^\mathrm{s}$ and $\mathrm{T}\mbox{-}K\alpha^\mathrm{hs}$, originating from single- and double $K$-shell vacancy production, respectively. Note, that the $\mathrm{T}\mbox{-}K\alpha^\mathrm{s}$ also includes photons from the second decay following a double $K$-shell vacancy production. 

The intensity of the projectile radiation is significantly larger than that of the target radiation, indicating that electron transfer from target to projectile has a much larger probability for $n\geq2$ levels than for the  $K$-shell--to--$K$-shell levels. Unfortunately, the $\mathrm{T}\mbox{-}K\beta^\mathrm{s,hs}$ lines are partially blended with the $\mathrm{P}\mbox{-}K\alpha$ lines and thus not visible.

The low intensity of the $\mathrm{T}\mbox{-}K\alpha^\mathrm{s,hs}$ radiation in relation to the $\mathrm{P}\mbox{-}K\alpha$ radiation is one of the challenges of these measurements. In addition, the effective solid angle for projectile radiation is enhanced due to the Doppler boost at forward detection angles. When correcting for this kinematic effect, our data show that the ratio of $\mathrm{T}\mbox{-}K\alpha^\mathrm{s,hs}$ to $\mathrm{P}\mbox{-}K\alpha$ radiation emitted at $35^\circ$ is $28\%$, $7\%$, and $2\%$ for 50, 30, and 15~MeV/u, respectively. In combination with the reduced energetic separation between target and projectile radiation towards lower collision energies, this situation constitutes a major challenge for measuring the $\mathrm{T}\mbox{-}K\alpha^\mathrm{s,hs}$ radiation with sufficient statistics.

\subsection{Target x-rays}

\begin{figure}
	\centering
	\includegraphics[width=1\columnwidth]{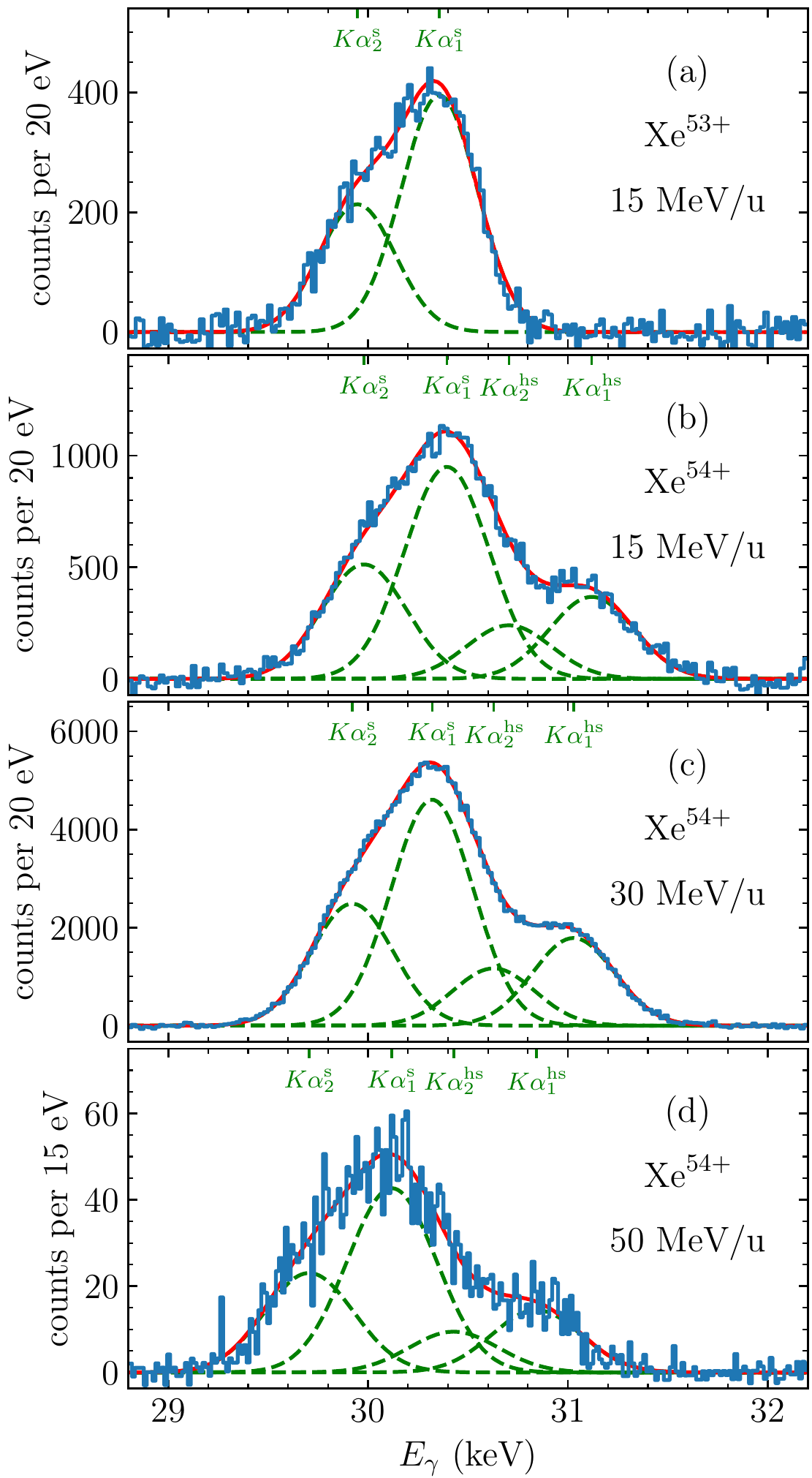}
	\caption{\label{fig:xraytarget}Background- and efficiency-corrected x-ray spectra of target $K\alpha_{2,1}^\mathrm{s, hs}$ radiation. The charge and energy of the projectile is given in the respective label. The measured spectra (blue lines) were fitted with Eqs.~(\ref{eq:fit}-\ref{eq:intensities}) (red lines). The centroid position of the individual fitted lines (dashed green lines) are marked on the corresponding top axes.}
\end{figure}

In Fig.~\ref{fig:xraytarget}(a), the target $K\alpha_{2,1}^\mathrm{s}$ radiation is shown for Xe$^{53+}(1s)$ projectiles. In this system, double electron transfer from the target $K$-shell to the projectile $K$-shell is highly improbable. The absence of the target $K\alpha_{2,1}^\mathrm{hs}$ radiation indicates that the probability for creating a target double $K$-shell vacancy through the collision with a Xe$^{53+}(1s)$ projectile is negligible since here only one electron can be captured from the target $K$-shell into the projectile $K$-shell, while the other $K$-shell  electron of the target would have to be captured into an excited state of the projectile or ionized into the continuum in order to create a target double $K$-shell vacancy. In Figs.~\ref{fig:xraytarget}(b)-\ref{fig:xraytarget}(d), the target $K\alpha_{2,1}^\mathrm{s,hs}$ radiation measured for Xe$^{54+}$ projectiles with an empty $K$-shell at three collision energies is shown. Figure~\ref{fig:xraytarget}(d) is based on the experimental data of Ref.~\cite{gumberidze_impact_2017}, which unfortunately contains poor statistics. The comparison of the x-ray spectra for the H-like Xe$^{53+}$ and bare Xe$^{54+}$ projectiles, and the absence of the $K\alpha_{2,1}^\mathrm{hs}$ radiation for the Xe$^{53+}$ projectiles indicate, that at low collision energies a target double $K$-shell vacancy is created by resonant transfer of two target $K$-shell electrons into an \textit{empty} $K$-shell of the projectile.

\begin{table*}[t!]
	\caption{\label{tab:results} Comparison of our experimental results with theory for target single and double $K$-shell vacancy production. Experimental errors reflect uncertainties for the fit parameters of Eqs.~(\ref{eq:fit}) and (\ref{eq:intensities}).}
	\begin{ruledtabular}
		\renewcommand{\arraystretch}{1.2}
		\begin{tabular}{lcdlllllrrd}
			& & & \multicolumn{5}{c}{Fit to experimental data}  & \multicolumn{3}{c}{Theory}\\
			\cline{4-8} \cline{9-11}
			Projectile & $E_\text{p}$ & \multicolumn{1}{c}{$v_K/v_\text{p}$} & \multicolumn{1}{c}{$E_{K\alpha_2^\mathrm{s}}$} & \multicolumn{1}{c}{$E_{K\alpha_1^\mathrm{s}}$} & \multicolumn{1}{c}{$\Delta E_\mathrm{hs}$} & \multicolumn{1}{c}{$\Delta E_\mathrm{FWHM}$} & \multicolumn{1}{c}{$\sigma_{K^{\text{-}2}}/\sigma_{K^{\text{-}1}}$} & \multicolumn{1}{c}{$\sigma_{K^{\text{-}1}}$} & \multicolumn{1}{c}{$\sigma_{K^{\text{-}2}}$} &
			\multicolumn{1}{c}{$\sigma_{K^{\text{-}2}}/\sigma_{K^{\text{-}1}}$}\\
			\cline{4-7} \cline{9-10}
			& \multicolumn{1}{c}{(MeV/u)} &  & \multicolumn{4}{c}{(eV)} & \multicolumn{1}{c}{(\%)} & \multicolumn{2}{c}{(kbarn)} & \multicolumn{1}{c}{(\%)}\\
			\hline
			Xe$^{53+}$ & 15 & 2.20   & $29946\pm7$ & $30356\pm3$  & --- & $446\pm9$ & --- & 325 & 9 & 2.8\\
			Xe$^{54+}$ & 15 & 2.20  & $29979\pm7$  & $30394\pm4$  & $724\pm8$ & $498\pm8$ & $71.1\pm3.5$ & 252 & 198 & 78.6\\
			Xe$^{54+}$ & 30 & 1.55  & $29920\pm2$  & $30320\pm1$  & $707\pm3$  & $482\pm3$ & $71.4\pm1.1$ & 199 & 154 & 77.4\\
			Xe$^{54+}$ & 50 & 1.20 & $29705\pm17$ & $30118\pm9$ & $723\pm23$ & $526\pm22$\footnote{The x-ray spectrum of Ref.~\cite{gumberidze_impact_2017} had a slightly worse energy resolution.} & $57.3\pm6.8$ & 192 & 97 & 50.5\\
		\end{tabular}
	\end{ruledtabular}
\end{table*}

Each x-ray spectrum was corrected for the energy-dependency of the detection efficiency and fitted by a superposition of four Gaussians,
\begin{equation}\label{eq:fit}
	I(E_\gamma)=a_0 + a_1 E_\gamma + \sum_{i=1}^{4} I_i~\mathrm{exp} \left(-\frac{(E_\gamma-E_i)^2}{2\sigma_\gamma^2} \right),
\end{equation}
corresponding to the four expected peaks, \mbox{$i\in\{K\alpha_2^\mathrm{s},K\alpha_1^\mathrm{s},K\alpha_2^\mathrm{hs},K\alpha_1^\mathrm{hs}\}$}. The coefficients $a_0$ and $a_1$ describe a linear background, which has been subtracted in the spectra shown in Fig.~\ref{fig:xraytarget}. The centroid energies were fitted with
$E_{K\alpha_2^\mathrm{hs}} = E_{K\alpha_2^\mathrm{s}} + \Delta E_\mathrm{hs}$ and 
$E_{K\alpha_1^\mathrm{hs}} = E_{K\alpha_1^\mathrm{s}} + \Delta E_\mathrm{hs}$, implying that the fine structure splitting of the satellite and the hypersatellite line are identical on the level of accuracy relevant here. (Deviations from this assumption are up to 1\%, see Sec.~\ref{sec:discussion}). The width of each peak was considered to be identical, characterized by the standard deviation $\sigma_\gamma$, with $\Delta E_\mathrm{FWHM}=2.36\sigma_\gamma$. The line intensities, $I_i$, of the fit function (\ref{eq:fit}) are given by
\begin{subequations}\label{eq:intensities}
	\begin{eqnarray}
		I_{K\alpha_2^\mathrm{s}} & = & I_0  \left( 1+ \frac{\sigma_{K^{\text{-}2}}}{\sigma_{K^{\text{-}1}}} \right) \omega_{K\alpha_2^\mathrm{s}}, \label{eq:int1}\\
		I_{K\alpha_1^\mathrm{s}} & = & I_0  \left( 1+\frac{\sigma_{K^{\text{-}2}}}{\sigma_{K^{\text{-}1}}} \right) \omega_{K\alpha_1^\mathrm{s}}, \label{eq:int2}\\
		I_{K\alpha_2^\mathrm{hs}} & = & I_0  \frac{\sigma_{K^{\text{-}2}}}{\sigma_{K^{\text{-}1}}} \omega_{K\alpha_2^\mathrm{hs}}, \label{eq:int3}\\
		I_{K\alpha_1^\mathrm{hs}} & = & I_0  \frac{\sigma_{K^{\text{-}2}}}{\sigma_{K^{\text{-}1}}} \omega_{K\alpha_1^\mathrm{hs}}. \label{eq:int4}
	\end{eqnarray}
\end{subequations}
Eqs.~(\ref{eq:int1}) and (\ref{eq:int2}) take into account that each hypersatellite photon is most likely followed by a satellite photon. The relative branching fractions of the radiative decay channels, $\omega_{K\alpha_2^\mathrm{s}}=0.3506\pm0.0002$ \cite{scofield_relativistic_1974} and $\omega_{K\alpha_2^\mathrm{hs}}=0.3953\pm0.0008$ \cite{costa_relativistic_2007}, were taken from theory, with $\omega_{K\alpha_1^\mathrm{s}}\equiv 1-\omega_{K\alpha_2^\mathrm{s}}$ and $\omega_{K\alpha_1^\mathrm{hs}}\equiv 1-\omega_{K\alpha_2^\mathrm{hs}}$. Another fit parameter is the overall intensity $I_0$. By using this fit model, we experimentally derived the cross section ratios  $\sigma_{K^{\text{-}2}}/\sigma_{K^{\text{-}1}}$ for the three investigated collision energies.

The fit results are summarized in Tab.~\ref{tab:results}. All errors refer to fit uncertainties at one standard deviation. The energy calibration uncertainty is about $1$~eV. Uncertainties in the detection efficiency correction over the energy range of the target $K\alpha_{2,1}^\mathrm{s,hs}$ radiation are negligible. The overall uncertainty of $\sigma_{K^{\text{-}2}}/\sigma_{K^{\text{-}1}}$ is discussed in Sec.~\ref{sec:discussion}.

\section{Theory}\label{sec:theory}

\begin{figure*}
	\subfigure{\includegraphics[width=1\columnwidth]{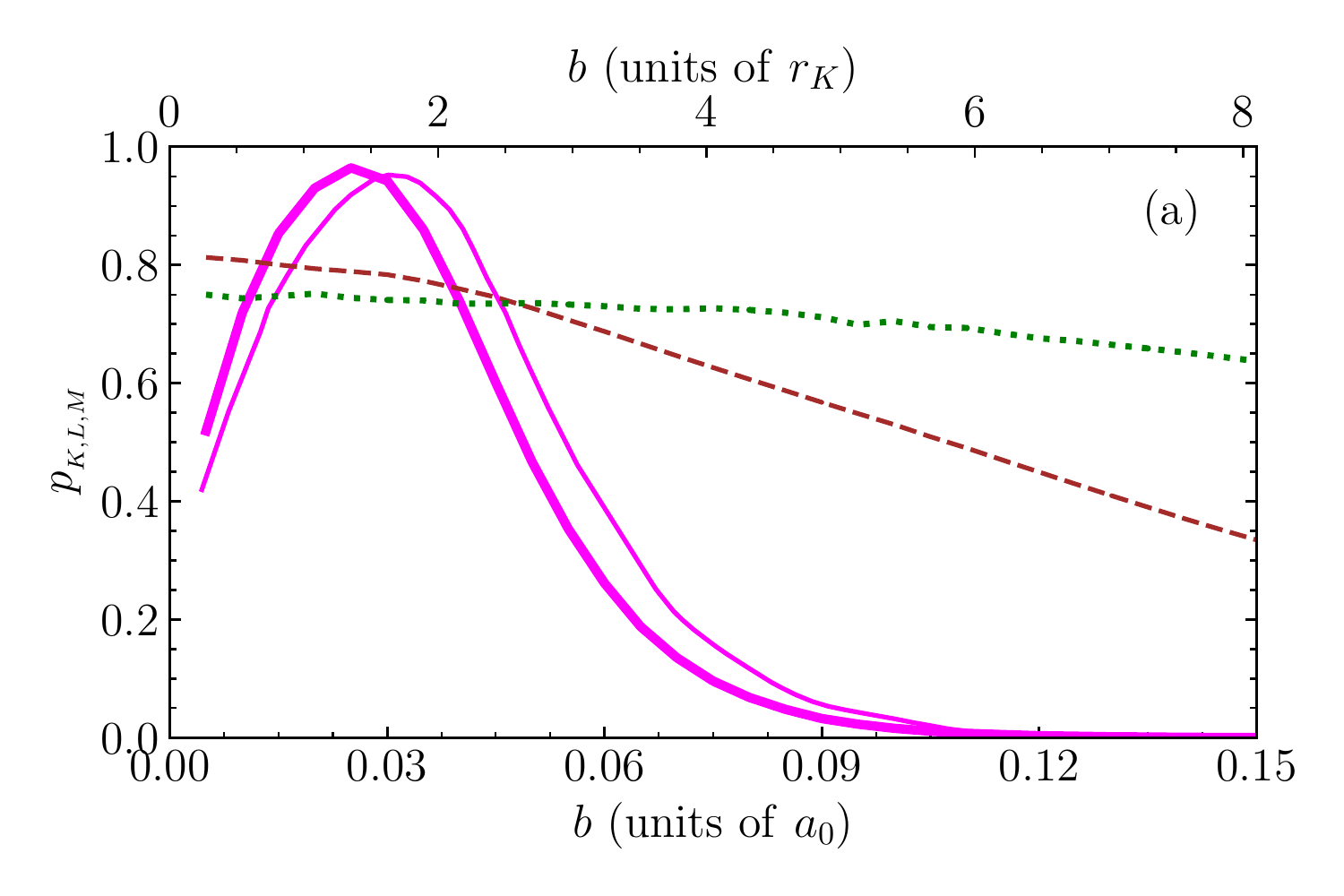}}
	\subfigure{\includegraphics[width=1\columnwidth]{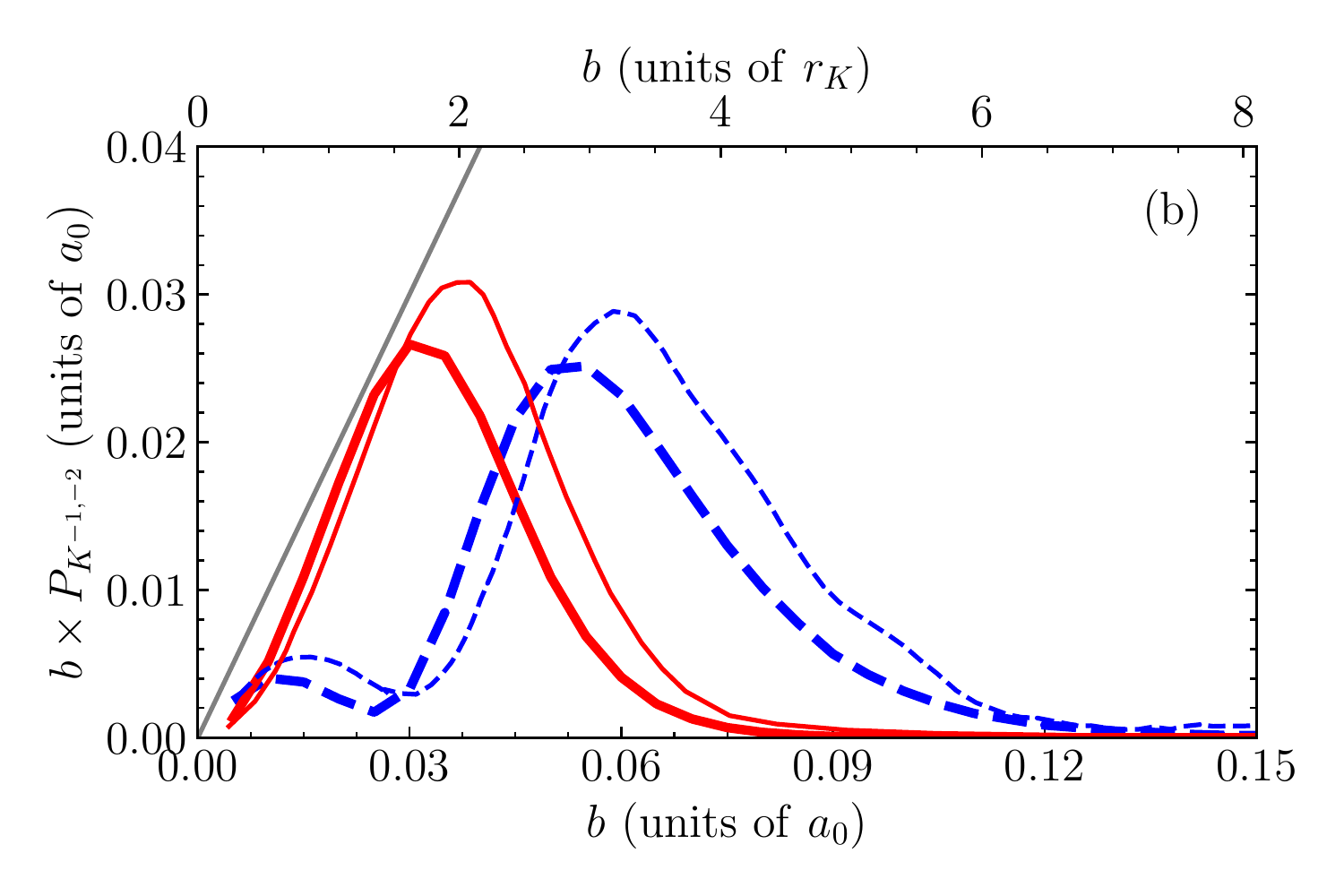}}
	\caption{\label{fig:theo}Theoretical results for Xe$^{54+}$ + Xe at 30~MeV/u, partially taken from Ref.~\cite{kozhedub_non-perturbative_2015}. Thick and thin lines show the relativistic (rel.) and the nonrelativistic (nonrel.) results, respectively. (a) Probability for an individual  target vacancy production in the $K$-shell (rel.: thick magenta line, nonrel.: thin magneta line), $L$-shell (rel.: dashed brown line), and $M$-shell (rel.: dotted green line). (b) Weighted probabilities for single (dashed blue lines) and double (solid red lines) target $K$-shell vacancy production. The gray diagonal line represents $P(b)=1$.}
\end{figure*}

The fully-relativistic time-dependent two-center approach applied in the present work was developed in Refs.~\cite{tupitsyn_relativistic_2010,tupitsyn_relativistic_2012} and has been compared to experimental results in Refs.~\cite{kozhedub_relativistic_2014,kozhedub_intensities_2017,schuch_quantum_2020}. For Xe$^{54+}+$Xe collisions at 30~MeV/u, theoretical results for the impact-parameter dependent single and double $K$-shell vacancy production probabilities were presented in Ref.~\cite{kozhedub_non-perturbative_2015}. For the present analysis we extended the previous calculations to 15 and 50~MeV/u, also including $L$- and $M$-shell vacancy production probabilities.

In the following, $p_K(b)$, $p_L(b)$, and $p_M(b)$ denote the probabilities as function of the impact parameter, $b$, of producing an \textit{individual} vacancy in the respective shell of the neutral target atom during the collision with a bare projectile. The example results for 30~MeV/u are shown in Fig.~\ref{fig:theo}(a). It can be seen that $p_K(b)$ peaks close to the so-called chemical distance of $b\approx 2 r_K$, where $r_K$ is the classical $K$-shell radius in Xe.
For comparison, the nonrelativistic approximation, derived by setting the speed of light to infinity, leads to a $p_K(b)$ which is slightly shifted towards larger values of $b$. This feature can be seen as the signature for the relativistic shrinking of the 1$s$-wavefunction -- or in the quasi-molecular picture the $1s\sigma$ molecular orbit -- of the Xe-Xe system.

Statistically, the probabilities for creating certain numbers of vacancies within an individual fully occupied shell follow a binomial distribution. For the $K$-shell, this binomial distribution leads to the impact-parameter dependent probability for creating one or two vacancies in the $K$-shell, 
\begin{eqnarray}\label{eq:Pb}
	P_{K^\text{-1}}(b) &=& 2 p_K(b) \left[ 1-p_K(b) \right], \nonumber \\ 
	P_{K^\text{-2}}(b) &=& \left[ p_K(b) \right]^2,
\end{eqnarray}
respectively. Note, that $P_{K^\text{-1}}(b)\leq0.5$ for all possible values of $p_K(b)\in[0,1]$, and $P_{K^\text{-2}}(b) \geq P_{K^\text{-1}}(b)$ for $p_K(b)\in [0.67,1]$. Deviations from this statistical independent-particle picture may result from correlation effects beyond the one-electron approximation, which are considered here to be negligible due to the strong Coulomb-potential of the heavy nuclei. 

Figure~\ref{fig:theo}(b) shows the probabilities weighted by $b$.  As intuitively expected, $b\times P_{K^{\text{-}2}}(b)$ dominates at smaller $b$, while $b\times P_{K^{\text{-}1}}(b)$ dominates at larger $b$. The total cross sections for single and double $K$-shell vacancy production are then given by
\begin{equation}
	\sigma_{K^{\text{-}1,\text{-}2}}=2\pi  \int_0^\infty b\times P_{K^{\text{-}1,\text{-}2}}(b)~db.
\end{equation}
The probabilities in the relativistic calculations peak at slightly smaller values of $b$ than in the nonrelativistic case, leading to slightly smaller cross sections. Note, that the differences between the relativistic and the nonrelativistic case essentially cancel out in the cross section ratio $\sigma_{K^{\text{-}2}}/\sigma_{K^{\text{-}1}}$. 

Within the range of impact parameters relevant for $p_K(b)$, the probabilities $p_L(b)$ and $p_M(b)$ are nearly constant. In the example shown in Fig.~\ref{fig:theo}, the Xe target atom loses in average 5.5 out of 8 $L$-shell electrons and 13.1 out of 18 $M$-shell electrons simultaneously with a single $K$-shell vacancy production, or 6.1 $L$-shell and 13.3 $M$-shell electrons simultaneously with double $K$-shell vacancy production. Similar numbers apply to the initially fully occupied $N$- and $O$-shell.

Furthermore, we illustrate that the target $K$-shell vacancy production mainly proceeds via $K$-$K$ charge transfer to the projectile by the following numbers: The double $K$-shell vacancy production in the target atom leads to a double $K$-shell population in the projectile ion with a probability of 95\%, 81\%, and 66\% for 15, 30, and 50~MeV/u, respectively. We note that it is not possible to unambiguously identify these events experimentally through a coincidence detection of the target $K\alpha_{2,1}^\mathrm{s,hs}$ photon and the down-charged projectile due to the large probability of simultaneous $n\geq2$ charge transfer from the target to the projectile.

\section{Discussion}\label{sec:discussion}

\begin{table}[b!]
	\caption{\label{tab:energies} Energies in eV for $E1$-transitions of different xenon charge states (rounded for easier readability).}
	\begin{ruledtabular}
		\renewcommand{\arraystretch}{1.2}
		\begin{tabular}{llccr}
			Target & Configuration & $K\alpha_2$ & $K\alpha_1$ & Ref.\\
			\hline
			Xe$^+$ & $1s^{\text{-}1}$ & 29458 & 29778 & \cite{deslattes_x-ray_2003}\\
			Xe$^{2+}$ & $1s^{\text{-}2}$ & 30093 & 30418 & \cite{costa_relativistic_2007}\\
			Xe$^{52+}$ & $1s2p$ & 30206 & 30630 & \cite{kozhedub_qed_2019} \\	
			Xe$^{53+}$ & $2p$ & 30857 & 31284 & \cite{yerokhin_lamb_2015} \\	
		\end{tabular}
    \end{ruledtabular}	
\end{table}

The positions of the experimentally derived line centroids, $E_{K\alpha_2^\mathrm{s}}$ and $E_{K\alpha_1^\mathrm{s}}$, given in Tab.~\ref{tab:results}, can be compared to the transition energies of different charge states of the xenon target presented in Tab.~\ref{tab:energies}. Note that the x-ray spectra of satellite and hypersatellite lines with all combinations of $L$-shell populations, but completely filled outer shells, have been calculated e.g.~in Ref.~\cite{chen_systematic_2015}. The values derived in our experiment for $E_{K\alpha_2^\mathrm{s}}$ and $E_{K\alpha_1^\mathrm{s}}$ lie consistently between the corresponding energies of the Xe$^+(1s^{\text{-}1})$ and the Xe$^{52+}(1s2p)$ configuration, while the values of $E_{K\alpha_2^\mathrm{hs}}=E_{K\alpha_2^\mathrm{s}}+\Delta E_\mathrm{hs}$ and $E_{K\alpha_1^\mathrm{hs}}=E_{K\alpha_1^\mathrm{s}}+\Delta E_\mathrm{hs}$ lie consistently between the corresponding energies of the Xe$^{2+}(1s^{\text{-}2})$ and the Xe$^{53+}(2p)$ configuration. This shows that the initially neutral Xe target is highly ionized after the collision, while the degree of ionization increases with decreasing collision energy.  

The measured line widths, $\Delta E_\mathrm{FWHM}$, given in Tab.~\ref{tab:results} originate from a quadrature sum of the broad (hyper)satellite x-ray distribution due to the broad target charge-state distribution after the collision \cite{chen_systematic_2015}, and the detector resolution (see Sec.~\ref{sec:experiment}), with both contributions being comparable in magnitude. This means that even with an ideal detector resolution, the four lines would still be energetically overlapping.

In our collision systems, the relative branching fractions of the radiative decay channels, $\omega_{K\alpha_2^\mathrm{s}}$ and $\omega_{K\alpha_2^\mathrm{hs}}$, result from a superposition of various charge states, subshell populations, and decay channels, and may thus deviate from those given in the literature for ions with completely filled outer shells \cite{scofield_relativistic_1974,costa_relativistic_2007}. Furthermore, a single or double $K$-shell vacancy can not only decay through $K\alpha_{2,1}^\mathrm{s}$ and $K\alpha_{2,1}^\mathrm{hs}$ emission, but with a minor probability also through $K\beta_{3,1}$ emission or Auger decay. The assumption of the fit model is that the branching fractions for $K\beta_{3,1}$-radiation, as well as the Auger yield, are identical for the satellite and the hypersatellite configurations. 

\begin{figure}
	\centering
	\includegraphics[width=1\columnwidth]{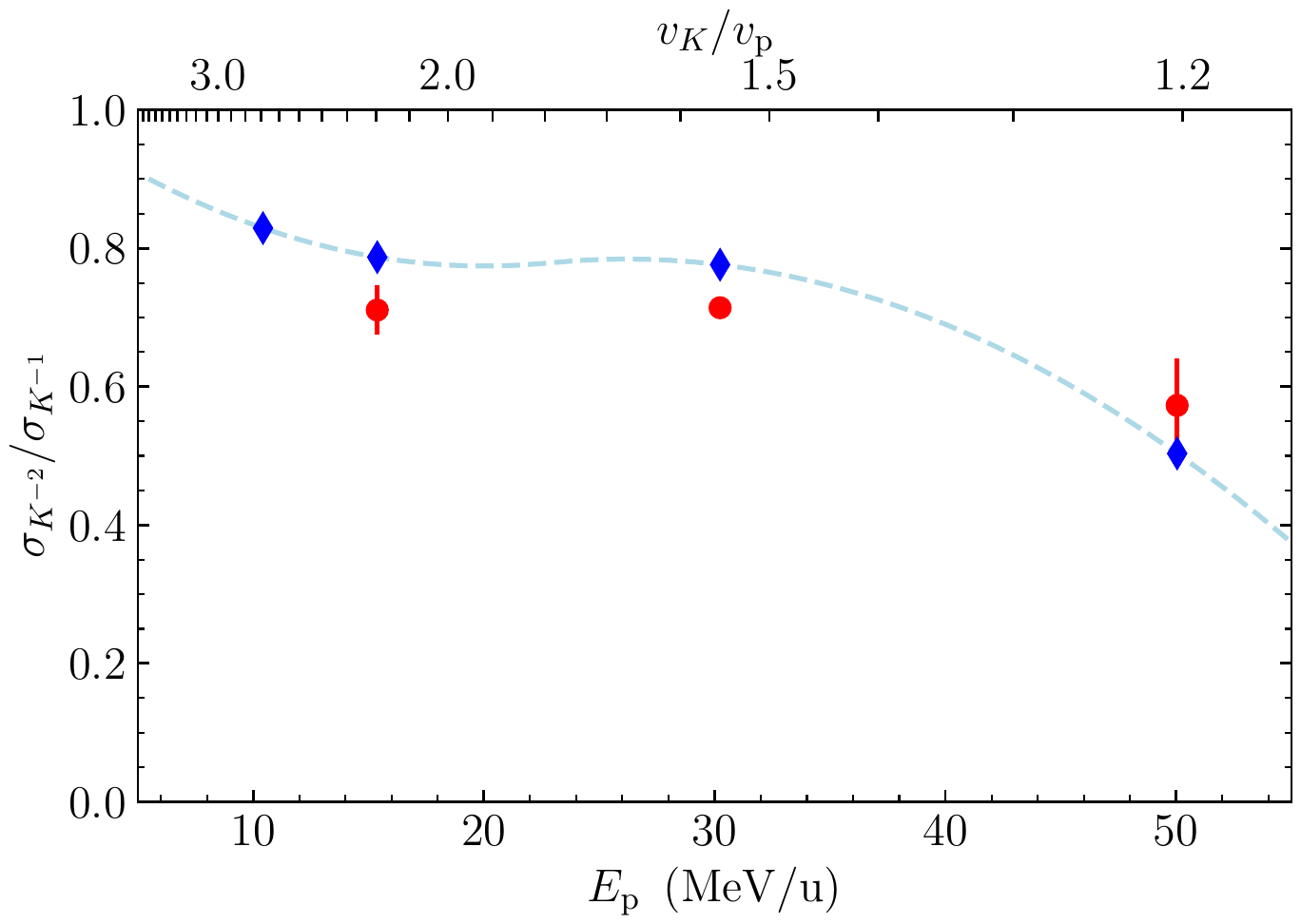}
	\caption{\label{fig:summary} Cross section ratio of double-to-single target $K$-shell vacancy production. Experimental results given in Tab.~\ref{tab:results} are shown as red points with statistical error bars. Theoretical results are shown as blue diamonds, connected by a dashed line as a guide to the eye.}
\end{figure}

In Fig.~\ref{fig:summary} we compare our experimentally determined ratio for the double-to-single target $K$-shell vacancy production cross section to the theoretical results. A reasonable agreement is evident. Presently, we cannot conclusively identify whether the remaining discrepancy originates from inaccuracies in the theoretical description of the process or in the potentially incomplete fit model we used to describe our data. From the theoretical side, electron correlation effects between the $K$-shell electrons, that were neglected due to the much stronger interaction of the electrons with the nuclei, may affect the cross sections by up to a few percent. In order to estimate this effect we performed a set of calculations using different kinds of the atomic screening potential. Also, numerical uncertainties may play a role, whose estimation is difficult. On the experimental side, measuring the target radiation with higher energy resolution using a microcalorimeter would enable us to apply an unconstrained fit to the four peaks and would thus reduce the assumptions required to describe the data by a fit model \cite{hengstler_towards_2015}.

\section{Summary and Outlook}

In summary, we experimentally studied slow collisions of bare and H-like xenon ions with neutral xenon target atoms. We performed x-ray spectroscopy of the target and projectile radiation emitted under a forward angle of $35^\circ$ with respect to the projectile beam direction. Our experiment signifies a major step within the regime of measurements at strong perturbations and provides some technical lessons learned for the path towards studying symmetric quasi-molecular collisions at even lower collision energies and supercritical fields. Most importantly, the intensity 
of the $K\alpha^\mathrm{s,hs}$ radiation relative to the projectile $K\alpha$ radiation rapidly decreases when going to lower collision energies, thus leading to a diminishing contribution of target $K$-shell vacancy production relative to charge transfer between outer shells.  This poses the main challenge when performing measurements at even lower collision energies. In addition, the Doppler shift decreases, which requires a better resolution and/or a smaller observation angle with respect to the projectile beam for energetically separating the target and projectile radiation.

In the presented study, we analyzed the target x-ray spectra to derive cross section ratios for double-to-single $K$-shell vacancy production for projectile energies ranging from 15 to 50~MeV/u. We compared these cross section ratios to the predictions of relativistic time-dependent two-center theory. Considering the assumptions applied in the analysis, we find a fairly good agreement. In a future study, the assumptions introduced in the analysis can be reduced significantly by enhancing the x-ray energy resolution using high-resolution magnetic microcalorimeters \cite{hengstler_towards_2015}, thus providing a more stringent test to theory. Corresponding experiments at collision energies down to a few MeV/u are now accessible at CRYRING@ESR, a dedicated low-energy storage ring, CRYRING, installed behind ESR \cite{lestinsky_physics_2016}.

\begin{acknowledgments}
We thank M.~Steck, S.~Litvinov, B.~Lorentz, and R.~Heß for operating the ESR. This research has been conducted in the framework of the SPARC collaboration, experiment E132 of FAIR Phase-0 supported by GSI. It is further supported by the European Research Council (ERC) under the European Union's Horizon 2020 research and innovation programme, grant No.~682841 ("ASTRUm") and the grant agreement No.~654002 (ENSAR2). We acknowledge substantial support by ErUM-FSP APPA (BMBF No.~05P15RGFAA and No.~05P19SJFAA) and by the State of Hesse within the Research Cluster ELEMENTS (Project ID 500/10.006). And E.P.B., S.N., and H.R. gratefully acknowledge support by the “Transnational Access to GSI” (TNA) activity.
\end{acknowledgments}

\bibliography{paperXenon}

\end{document}